\documentclass[aps,prd,showpacs,floatfix,nofootinbib,twocolumn,10pt]{revtex4-1}

\usepackage{color,amsmath,amssymb,graphicx,latexsym,subfigure}
\usepackage{soul,color}
\usepackage[dvipsnames]{xcolor}

\usepackage{tabularx} 
\usepackage{amsmath}  
\usepackage{graphicx} 
\usepackage[final]{hyperref} 
\hypersetup{
	colorlinks=true,       
	linkcolor=blue,        
	citecolor=blue,        
	filecolor=magenta,     
	urlcolor=blue         
}

\usepackage{ezedits}

\newcommand{\beq}{\begin{equation}}
\newcommand{\eeq}{\end{equation}}
\newcommand{\bea}{\begin{eqnarray}}
\newcommand{\eea}{\end{eqnarray}}

\def\toff{t_\text{on}}

\begin{document}

\title{Searching for axion streams with the echo method}

\author{Ariel Arza$^{a,b}$, Abaz Kryemadhi$^{c}$ and Konstantin Zioutas$^{d}$
}
\affiliation{
$^{a}$ Tsung-Dao Lee Institute (TDLI), Shanghai Jiao Tong University, Shanghai 200240, China
\\
$^{b}$ Institute for Theoretical and Mathematical Physics (ITMP),
Lomonosov Moscow State University, 119991 Moscow, Russia
\\
$^{c}$ Dept. Computing,Math \& Physics, Messiah University, Mechanicsburg PA 17055, USA
\\
$^{d}$ Physics Department, University of Patras, GR 26504,Patras, Greece
}

\begin{abstract}

The axion dark matter echo technique, proposed in Ref. \cite{Arza:2019nta}, aims to search for axion dark matter by detecting the electromagnetic echo coming from stimulated decay of ambient axion dark matter interacting with a radio-microwave outgoing beam. In this work we consider deviations from the standard halo model (SHM) in the form of fine grained streams that are present in the solar system and use them as a target for the axion echo method, which demonstrates to be very sensitive to particular local axion dark matter phase space distributions. We show that the extremely small dispersion of the streams works in favor of the echo method performance, improving its sensitivity in the axion-photon coupling up to two orders of magnitude, with respect to the isotropic SHM. We also discuss the possibility of targeting high density dark matter hairs formed when fine grained streams focus by the inner Earth's gravitational field.

\end{abstract}

\maketitle

\section{Introduction}

A number of astrophysical and cosmological observations establish that baryonic matter constitutes  only a $\sim4\%$ of the total energy budget of the universe. Most of the energy density, an amount of $\sim70\%$, is attributable to dark energy, and the remaining of about $\sim26\%$ is assumed to be dark matter (DM), a non-relativistic matter component that is essential to explain the evolution of the universe and its structure formation. So far, direct DM searches have not shown any interaction with the visible matter other than gravity. Thus, from the particle physics point of view, its identity remains an open question. In the particle hypothesis for the dark matter composition, among a variety of candidates such as weakly interacting massive particles (WIMPs) \cite{Arcadi:2017kky}, dark photons \cite{Jaeckel:2012mjv}, sterile neutrinos \cite{Boyarsky:2018tvu}, millicharged particles \cite{Bogorad:2021uew}, etc., axions and also axion-like \cite{Arias:2012az} particles have reached great attention in the last decade.

The QCD axion is among the best motivated candidates for the invisible Universe we are living in. It is theoretically inspired following the Peccei-Quinn mechanism \cite{Peccei:1977hh}. Its most prominent  manifestation is the solution of the strong CP problem, namely, the puzzle why the strong interactions conserve P and CP \cite{Wilczek:1977pj,Weinberg:1977ma}. On the other hand, invisible axion models \cite{Kim:1979if,Shifman:1979if,Dine:1981rt,Zhitnitsky:1980tq} are very attractive from the cosmological point of view since axions can be produced in the early universe by non thermal means, leading to a cold and abundant population to fit the today's dark matter energy density \cite{Preskill:1982cy,Abbott:1982af,Dine:1982ah}. Apart from the QCD axion, different extensions of the standard model also predict the existence of axion-like particles, for instance in any string compactification \cite{Svrcek:2006yi}. These axion-like particles are phenomenologically similar to the QCD axion, with the difference that its mass and couplings to the standard model particles are not related to each other, which does occur for the QCD axion. 

Direct searches for axion dark matter consist mainly in detecting electromagnetic signals coming from the interaction between the axion dark matter field and strong magnetic fields \cite{Sikivie:1983ip}. Resonant cavity experiments are the most developed axion dark matter techniques, reaching sensitivities compatible with parameter space of QCD axion models \cite{ADMX:2018gho,ADMX:2019uok,ADMX:2021nhd, Adair:2022rtw}. There are also a number of other proposals aiming to do so in the future (see Ref’s \cite{Beurthey:2020yuq,DMRadio:2022pkf}), and several other different setups and ideas \cite{Sikivie:2020zpn,Irastorza:2018dyq}. One new idea is the echo method considered here \cite{Arza:2019nta}, which is based on the stimulated decay of local DM axions into photons. It aims to send in outer space a powerful beam of electromagnetic radiation, in the radio-microwave band. The high photon occupation number of the beam results to a Bose enhancement effect on the axion to two photons decay mode. Following this forced axion decay, because of momentum conservation, one of the produced photons adds to the outgoing beam while the other one is released in the opposite direction to the outgoing beam propagation, as an echo signature. The plan of the method is to detect this echo wave that travels backwards with respect to the outgoing beam propagation. Energy conservation requires that the stimulated axion decay occurs when the outgoing beam frequency is near half of the axion mass $m$, and the echo signal is a spectral line whose frequency is centered very close to $m/2$ as well, being slightly red (blue) shifted depending on the relative velocity of axions with respect to the lab frame.       

For most of DM experimental setups, the signals depend on the local DM phase space distribution. Although deviations from the SHM can be revealed from the signals \cite{Foster:2017hbq}, the effects are weak. This is, however, not the case for the echo method which is demonstrated to be very sensitive to certain properties of the dark matter phase space distribution. These effects were discussed in both the original proposal~\cite{Arza:2019nta} as well as the detailed signal calculation of Ref. \cite{Arza:2021nec}. The echo method can be very efficient for scenarios dominated by very cold flows, i.e., low velocity dispersion. In particular, it was shown in \cite{Arza:2019nta,Arza:2021nec} that the performance of the echo method is improved notably when considering the caustic ring model \cite{Duffy:2008dk,Chakrabarty:2020qgm}.

As the echo method is sensible to the axion phase space distribution, it provides an opportunity to search for axion dark matter in certain local dark matter models. One of these models predicts the existence of fine grained streams in the solar system~\cite{Vogelsberger:2011,Stucker:2020,Vogelsberger:2009,Vogelsberger:2008, Sikivie:1995,SikivieVelo:1999,natarajan:2008}. They are primordial cold dark matter streams with very small dispersion that are remnant from the early stages of galactic formation. Their existence is supported by N-body simulations \cite{Stucker:2020,Vogelsberger:2008,Vogelsberger:2009,Vogelsberger:2011,Vogelsberger:2020}. In fact, in Ref.~\cite{Vogelsberger:2011} it was shown that up to $10^{12}$ such streams could be at solar vicinity, and the average nominal dark matter density at a particular location is the superposition of the energy density of the various streams.  

This work studies the echo method performance when considering that the local dark matter distribution is composed by fine grained streams. As shown throughout the paper, the small velocity dispersion of the streams as well as their preferred directionality, make the echo method a very sensitive technique for axion dark matter search. Indeed, the projected sensitivity is up to two order of magnitude higher, when comparing with the smooth SHM. 

This work focuses on two scenarios. A) targeting low density fine grained streams by sending the outgoing electromagnetic beam in a random direction, and B) by sending the outgoing beam vertically in order to probe high density axion “hairs”, which are formed by gravitational self-focusing of streams passing through the Earth. Both scenarios have, a priori, some advantages and disadvantages. On the one hand, in the first case we have the advantage of targeting streams and receive echo as soon as the outgoing beam is turned on and regardless of the outgoing beam direction. However, the important disadvantage of this scenario is that the target has usually very low density. As shown below, high density encounters are rare. The second scenario looks very promising considering the high densities of the gravitationally derived hairs, of the order of $10\,\rho_0$ (where $\rho_0=0.45\,\text{GeV/cm}^3$ is the mean local DM energy density in the SHM) or even as high as $10^3\rho_0$. However, as the  directions in which the streams flow are unknown, the location of the hairs can not be predicted. Since a large number of hairs are expected, it is still possible for the simulations to estimate the encounter time scale that guarantees to observe at least one hair by looking at any direction in the sky. Despite the large hair density, the sensitivity in this case is limited by the observation time which is defined by the Earth angular velocity. To encounter a new stream, one would have to wait for the encounter time scale which, depending on the stream velocity, can vary from one day to one month.

The article is organized as follows. In Sec. \ref{streams} we introduce in more detail the concept of fine grained streams, its abundance and the gravitational focusing effect in the Earth. In Sec. \ref{echo signal} we describe the axion dark matter echo signal produced when a microwave-radio beam is sent out to space. In Sec. \ref{results} we show the sensitivity prospects for the echo method in presence of fine grained streams. It is realized in two scenarios. First, for the case where the photon beam is sent in a random direction and produces an axion echo from one of the low energy density ambient streams. Then, for the case when the beam encounters a stream hair produced by Earth's gravitational focusing. We finally conclude in Sec. \ref{conclusion}.
 
\section{Streams} \label{streams}

The standard halo model (SHM) assumes a smooth iso-thermal distribution of non-interacting dark matter particles with a Maxwellain velocity distribution. However, at the last scattering as the universe cooled one expects a tiny velocity dispersion for dark matter~\cite{Vogelsberger:2020,Stucker:2020}.  
Vogelsberger et al used a geodesic equation and an N-body simulation to find out that there are up to $10^{12}$ fine-grained streams at the solar system, where each stream carries a small fraction of the local DM density~\cite{Vogelsberger:2009,Vogelsberger:2011}. Interestingly for the purpose of the present work, the fine grained streams are expected to have a very low velocity dispersion, about $\sim 10^{-10}c$ ($c$ is the speed of light) for WIMPs and $\sim 10^{-17}c$ for axions~\cite{Vogelsberger:2009,Vogelsberger:2008, Sikivie:1995,SikivieVelo:1999,natarajan:2008}. The nominal DM density distribution at one point is described as a superposition of densities from all the individual streams~\cite{Vogelsberger:2011}, reproducing the standard DM profile.  

The stream abundance at Earth's position for a particular density ratio $\rho_{s}/\rho_{0}$ is estimated by using the Refs.~\cite{Vogelsberger:2011,gfstream:2022}. Table~\ref{tab:stream_table} lists the number of fine grained streams at solar neighborhood for specific density ratios. 
\begin{table}
	\centering
	\caption{Streams with different densities before gravitational focusing (GF) and for hairs after GF.}
	\label{tab:stream_table}
	\begin{tabular}{lcr} 
		\hline
		Density before GF & Number of streams & Density after GF\\
        ($\rho_s/\rho_{0}$) &   &  ($A \rho_s/\rho_{0}$) \\
		\hline
		1 & 1 & $10^8$\\
		0.1 & 1 & $10^7$\\
		0.01 & 1 & $10^6$ \\
		10$^{-3}$ & 10 & $10^5$\\
		10$^{-4}$ & 500 & $10^4$\\
		10$^{-5}$ & $2\cdot10^4$ & $10^3$ \\
        10$^{-6}$ & $4\cdot10^5$  & $10^2$ \\
        10$^{-7}$ & $2\cdot10^6$  & $10$\\
		\hline
	\end{tabular} 
\end{table}
 
\subsection{Gravitational focusing (GF) - axion hairs }
\begin{figure}[h!]
  \includegraphics[width=1\linewidth]{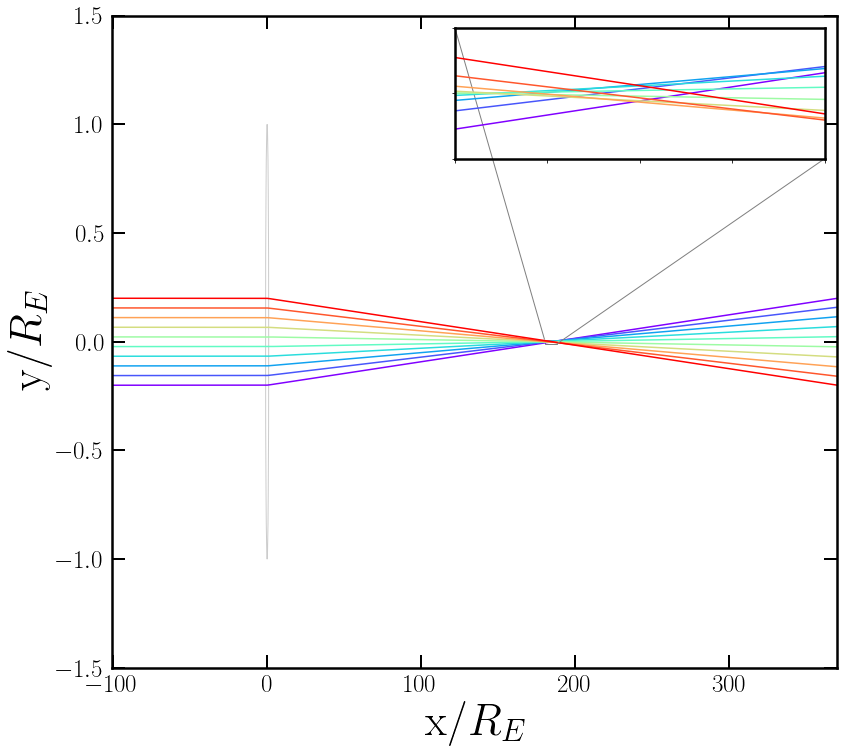}
\caption{A stream experiences gravitational effects of the Sun and then a self-focusing gravitational effect by the Earth. Gravitational self-focusing by the Earth from a stream with velocity of 220 km/s in the laboratory frame incoming from negative x and focusing the particles some distance away from the Earth \cite{Prezeau:2015}.} \label{fig:focusing}
\end{figure}

Once the streams reach the Earth they get focused on the opposite Earth's side due to the inner Earth's gravitational field, resulting in a small region of high density across the focus dubbed ``hair", as it is shown in Fig.~\ref{fig:focusing}. The amplification factor due to the self-focusing is effectively \cite{Prezeau:2015}
\beq
A=\frac{\pi r_{\text aperture}^2}{\pi r_{\text focus}^2}.
\eeq

The simulation of the present work uses the code of Ref.~\cite{gfstream:2022}, without taking into account the rest of the solar system. The amplification factor peaks at about $10^9$ for axion streams, being nearly velocity independent. The present simulations show that the dense region around the focus area can be modeled by a cylindrical shape with a length of $\sim 0.2 R_E$ ($R_E$ is the Earth's radius), a radius of $\sim 2.5\,\text{km}$, and with an average amplification of $10^8$. The location of a hair varies with the velocity of the incoming stream; thus the Earth acts for streams as a velocity selector. The distance of the focal point from the center of the Earth is given by~\cite{gfstream:2022, Prezeau:2015}
\beq
F \sim \frac{v^2}{(11.2\,\text{km/s})^2} R_{E}. \label{focalpoint}
\eeq

The density enhancement for one stream, due to GF by the inner Earth mass distribution, can be expressed as~\cite{gfstream:2022}
\beq
\rho \sim A \rho_{s},
\label{eq:densitygfunique}
\eeq
which applies even for a multi stream environment (for more details see Ref.~\cite{gfstream:2022}).  
 The last column in Table~\ref{tab:stream_table} shows the densities of hairs which are dominated by a single stream due to the high gravitational amplification ($\sim 10^8$).  
 
The simulation in~\cite{gfstream:2022} is used also to determine the abundance of hairs near the Earth as a function of observation distance $d_{\text{obs}}$ (distance between the root of the hair and the Earth's center) and the results are given in Table ~\ref{tab:daily_rate}.  

For a single radio telescope the daily rate of hairs is calculated for three different observation distances  $d_{obs}\leq 3 R_E$, $d_{obs}\leq 10 R_E$, and $d_{obs}\leq 20 R_E$~\cite{gfstream:2022}. 
\begin{table}
	\centering
	\caption{The daily rate of hairs under different conditions after GF.}
	\label{tab:daily_rate}
	\begin{tabular}{|l|c|r|r|} 
		\hline
		 $d_{obs}$ & $\frac{N_e}{day}$ & $\frac{N_e}{day}$ & $\frac{N_e}{day}$\\
                 &        ($\rho = 10 \rho_0$) &      ($\rho = 100 \rho_0$) &    ($\rho = 1000 \rho_0$)     \\
		\hline
		$\leq 3R_E$ & 0.05 & 0.004 & 0.0002\\
		\hline
        $\leq 10R_E$ & 0.2 & 0.02 & 0.001\\
		\hline
        $\leq 20R_E$ & 0.4 & 0.03 & 0.002\\
		\hline
	\end{tabular} 
\end{table} 

A radio telescope located on Earth's surface moves with a speed of $\sim 0.5~{\textrm {km/s}} \cdot \cos(\phi)$ due to Earth's rotation around its axis. For a radio telescope with a fixed beam direction the beam's transition time through a hair at an observation distance $d_{\text{obs}}$ is
\beq
t_{enc} \sim \frac{(2\cdot r_{ave})}{ (d_{\text{obs}}/R_{E})\, 0.5\,(\text {km/s}) \cos (\phi) },
\eeq  
where $r_{ave}$ is the average radius of the gravitationally focused region, and $\phi$ is the latitude of the receiver. 

\begin{figure}[h!]
  \includegraphics[width=1\linewidth]{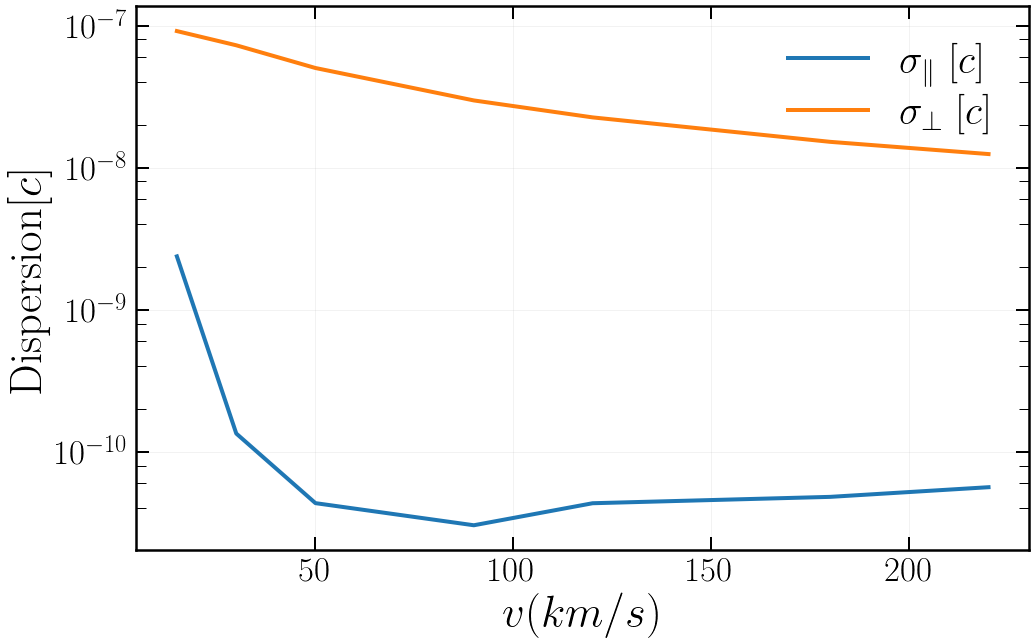}
\caption{Parallel and perpendicular dispersion of axion hairs. The decrease of dispersion at larger velocities is due to the fact that the larger the distance of the hair the more parallel the velocity component becomes.}
\label{fig:hairdispersion}
\end{figure}

Axions from incident streams, i.e., before they are affected by Earth's gravity, have an inherent dispersion of about $10^{-17}$c. However
, as the stream particles are deflected by the interior of the Earth, the velocity dispersion is increased by the gravitational deflection. Figure~\ref{fig:hairdispersion} gives the parallel and perpendicular velocity dispersion, they are about $\sim 10^{-10}c$ and $\sim 10^{-8}c$, respectively.  

For the echo method which depends not only on the density of streams but also on its dispersion we investigate  both cases of streams a) cosmological streams incident on the Earth but not yet affected by the inner Earth's gravitational field, and b) the high density hairs produced by the inner Earth's gravitational focusing of the streams which exit the Earth.   

\section{Echo signal} \label{echo signal}

We consider an electromagnetic beam propagating through an axion flow with extremely low velocity dispersion which is reasonable for the aforementioned fine grained streams. In this case the flow moves in some specific direction, then we define $v_\parallel$ and $v_\perp$ as the velocity components of the flow which are parallel (co-linear) and perpendicular to the beam, respectively. When the frequency of the photon beam is near half of the axion mass (see below), a significant echo signal is produced by stimulated axion decay. The echo is produced if the photons of the beam have an energy 
\beq
\omega={m\over2}(1+v_\parallel) \label{omega1}
\eeq
where $m$ is the axion mass. The echo photons have an energy
\beq
\Omega={m\over2}(1-v_\parallel) \label{Omega1}
\eeq
and are released in a direction that forms an angle of
\beq
\theta=2v_\perp = 6.7\times10^{-4}\,\text{rad}\left(v_\perp\over100\,\text{km/s}\right) \label{theta1}
\eeq
with respect to the photon beam direction. Notice that from Eqs.~\eqref{omega1} and \eqref{Omega1} it follows that for an outgoing beam propagating towards the same direction as the axion flow ($v_\parallel>0$), the echo frequency is red-shifted with respect to that of the outgoing beam, while in the opposite case ($v_\parallel<0$), the echo frequency is blue-shifted.

When the outgoing photon beam interacts with the axion flow over a spatial extent $l$, the expected total echo power is given by
\beq
P_\text{echo}={1\over16}g^2\rho{dP_0\over d\nu}l \enspace, \label{Pecho1}
\eeq
where $g$ is the axion-photon coupling, $\rho$ the axion dark matter energy density, and $dP_0\over d\nu$ the spectral power of the incident beam being evaluated at $\nu=\omega/2\pi$, with $\omega$ given by Eq. \eqref{omega1}.

This echo power signal is a very narrow spectral line with a bandwidth
\beq
B_s={m\delta v_\parallel\over4\pi} \enspace, \label{Bs1}
\eeq
where $\delta v_\parallel$ is the axion velocity dispersion of the component parallel to the photon beam. For axions, the velocity dispersion of the fine grained streams is of order $\delta v\sim10^{-17}c$, so the signal width might be as low as
\beq
B_s=1.21\times10^{-8}\,\text{Hz}\,\left({m\over10^{-5}\text{eV}}\right) \label{Bs2}
\eeq

The power given by Eq. \eqref{Pecho1} is the total power in the echo wave. Unfortunately, for most of the cases, this is not the power that can be collected by a receiver with limited size. The collected power is usually much smaller because the echo photons are released in the direction given by Eq. \eqref{theta1}, which makes the echo spread laterally in the direction of $\vec v_\perp$ (for more details, See Ref. \cite{Arza:2021nec}).

For an observation time $t_m$ and a total noise temperature $T_n$, the signal to noise ratio is given by Dicke's radiometer relation
\beq
s/n={P_c\over T_n}\sqrt{t_m\over B}, \label{s/n1}
\eeq
where $P_c$ is the total signal power collected by the receiver and $B$ is the frequency resolution of the observation.

In the following we address two scenarios:
a)  the target is any of the low density axion streams entering the solar system (See Tab. \ref{tab:stream_table}), which are assumed to be everywhere and interact with the outgoing beam all the time when it is on, and b) the target is a high density hair formed by stream focusing due to the Earth's gravitational potential. In this case, the target is not everywhere but at some particular location above the Earth surface with respect to the geo-center. Since their location is unknown, the outgoing photon beam for each measurement is turned on in a fixed direction for a time scale of the order of the encounter rate (see Sec. \ref{streams}). Of note, this is also dependent on the Earth's rotational velocity, the hair diameter, etc. 

\subsection{Low density axion streams} \label{echo signal low}

In the aforementioned scenario a),  the axion stream target is everywhere and interacts with the outgoing photon beam instantly as it is turned on. In the following, we assume that the photon beam is emitted from the geometrical center of a detector with effective radius $R_c$. As discussed in Ref.~\cite{Arza:2021nec}, for continuous outgoing beam emission, after a time $t\sim R_c/v_\perp$, the collected echo power in the detector reaches a constant value given by
\beq
P_c={1\over32\sqrt{2\pi}}{g^2\rho\over\Delta\nu}{R_c\over v_\perp}P_0 \enspace, \label{Pc1}
\eeq
assuming a Gaussian shape for the photon beam spectral power density with bandwidth $\Delta\nu$. Evidently, for small values of $v_\perp$ the echo wave is laterally less spread and more of the echo power is collected. As the direction of the stream is not known, the value of $v_\perp$ can not be arbitrarily small. If the stream direction is known with such a precision that $v_\perp$ is smaller than $v_\text{rot}$, which is the rotation velocity of the Earth surface, then the value of $v_\perp$ must be replaced by $v_\text{rot}$ in Eq.~\eqref{Pc1}.  

In the steady state described by Eq. \eqref{Pc1}, it is required that the outgoing beam is turned on for a time $\toff$ bigger than $R_c/v_\perp$. As fast as it is turned off, the echo signal drops abruptly to zero. So, the observation time for a single pulse is given essentially by $\toff$. The bandwidth of the signal is extremely narrow as it is given approximately by Eq. \eqref{Bs2}. However, the maximum frequency resolution is limited by the detector equipment. As a reference value, throughout the paper we assume a detector spectral line resolution of
\beq
B_d=1\text{Hz} \label{B1}
\eeq

which is achievable following modern channelization techniques \cite{Jiang:2019rnj}. For the low density axion stream scenario, Eq. \eqref{s/n1} is evaluated for $B=B_d$.

What is left is the value of $\toff$, which depends on the total time needed to scan a certain axion mass range. In this work we assume a total time period of $t_T=1\text{yr}$ to scan a factor of two in axion mass range. During this period of time, the number of outgoing photon pulses is $n=t_T/\toff$. The $n$ pulses are supposed to cover an octave in axion mass range. Then, the bandwidth of the outgoing beam for each shot is $\Delta\nu=m/(4\pi n)$, and $\toff$ is related to $\Delta\nu$ by
\beq
\toff={4\pi\Delta\nu\,t_T\over m}. \label{toffdeltaomega1}
\eeq
Notice that $2\pi\Delta\nu$ can not be arbitrarily small due to the uncertainty principle ($2\pi\Delta\nu\,\toff\gtrsim1$). Following Eq. \eqref{toffdeltaomega1}, the outgoing beam bandwidth is constrained to be larger than or equal to
\beq
\Delta\nu_\text{min}={1\over2\pi}\sqrt{m\over2t_T}\simeq2.47\,\text{Hz}\left({m\over10^{-5}\text{eV}}\right)^{1/2}\left({1\text{yr}\over t_T}\right)^{1/2}.
\eeq
Moreover, we quantify $\Delta\nu$ as being a fraction $\eta$ of the outgoing beam central frequency $\nu=m/(4\pi)$, which in terms of the axion mass is given by
\beq
\Delta\nu={\eta\,m\over4\pi} \label{Deltanu}
\eeq

Combining Eqs. \eqref{Pc1}, \eqref{toffdeltaomega1}, \eqref{Deltanu}, and the fact that $t_m=\toff$, the signal to noise ratio given in Eq. \eqref{s/n1} as a function of experimental parameters becomes
\beq
s/n={1\over8}\sqrt{\pi\over2}{g^2\rho\,P_0\over m\,T_n}{R_c\over v_\perp}\sqrt{t_T\over \eta B_d} \label{sn2}.
\eeq

\subsection{Axion hairs} \label{echo signal hairs}

To compute the echo signal when pointing at an axion hair, we use the results of the Appendix D in Ref. \cite{Arza:2021nec} to find the echo power at the base of the hair. The echo power in an area of radius $R_c$ is given by
\beq
P_c={1\over32\sqrt{2\pi}}{g^2\rho\over\Delta\nu}{\cal T}_cP_0 \label{Pc2}.
\eeq
The value of ${\cal T}_c$ depends on whether the length of the hair $l$ is bigger or smaller than $R_c/v_\perp$, which gives the relation
\beq
{\cal T}_c=\min\left(l,{R_c\over v_\perp}\right) \enspace. \label{Tc1}
\eeq

In this scenario the outgoing photon beam is assumed to propagate in a direction that is perpendicular to the Earth's surface.  Whenever the photon beam encounters a hair, they should be aligned. Then, the main contribution for $v_\perp$ comes from Earth's rotation velocity. We have roughly $v_\perp=v_r\cos(\phi)$, where $v_r$ is the velocity at the equator and $\phi$ is the latitude of the experiment's location. 

We recall that Eq. \eqref{Pc2} refers to the power at the base of the hair. Then, this is not necessarily the power collected at the Earth's surface where the receiver is located. Indeed, as the echo is emitted in the direction given by Eq. \eqref{theta1}, if the hair's root is located at a distance $h$ from the Earth's surface, the echo will be completely out of the detector area, if $R_c\leq2v_\perp h$. Therefore, the hair can not be located at heights farther than
\bea
h_\text{min} &=& {R_c\over 2v_\perp} \nonumber
\\
&=& 3.2\times10^4\,\text{km}\left({R_c\over 100\text{m}}\right)\left({434\text{m/s}\over v_\perp}\right) \enspace. \label{hmin1}
\eea
Following this constraint, the value of ${\cal T}_c$ is modified to
\beq
{\cal T}_c=\min\left(l,{R_c\over v_\perp}\right)
\begin{cases}
{R_c-2v_\perp h\over R_c}
& \quad\mathrm{for}\quad h<h_\text{min} 
\\
0
& \quad\mathrm{for}\quad  h>h_\text{min}  \enspace.
\end{cases}  \label{Tc1}
\eeq

Also, because of the Earth's rotation, the hair can only interact with the photon beam for a limited time interval. The observation time is given by the following equation:
\beq
t_m={D\over\omega_\text{rot} R_h} \label{tm1},
\eeq
where $D$ is the diameter of the hair, $R_h$ the position of the hair measured from the earth center, and $\omega_\text{rot}=2\pi/(24\text{hrs})$ is the angular frequency of the Earth's rotation. In terms of the stream velocity $v_s$ (see Eq. \eqref{focalpoint}), the observation time becomes
\beq
t_m=2.6\,\text{s}\left(D\over 6\text{km}\right)\left(25\text{km/s}\over v_s\right) \enspace. \label{tm2}
\eeq

With respect to the signal bandwidth, from the parallel velocity dispersion showed in Fig. \ref{fig:hairdispersion}, which is of the order $\delta v_\parallel\sim10^{-10}c$, we have
\beq
B_s={m\delta v_\parallel\over4\pi}=0.121\,\text{Hz}\left(m\over10^{-5}\text{eV}\right) \enspace. \label{Bs}
\eeq
This is barely smaller than our assumption made for Eq. \eqref{B1}. For this case we still assume $B_d=1\text{Hz}$ as a reference value for the detection capabilities.
We see that for axion masses smaller than $\sim10^{-4}\text{eV}$, $B_s$ is smaller than our assumed highest resolution given by $B_d=1\text{Hz}$ (see the discussion around Eq. \eqref{B1}). For this case, $B=1\text{Hz}$ is the value we must use when calculating the signal-to-noise ratio given by Eq. \eqref{s/n1}. However, for $m>10^{-4}\text{eV}$ we have $B_s>B_d$ and Eq. \eqref{s/n1} should be evaluated with $B=B_s$. Thus, $B$ becomes
\beq
B=\max\left(B_s,B_d\right) \label{B2}
\eeq

\begin{figure}[t]
\includegraphics[width=1\linewidth]{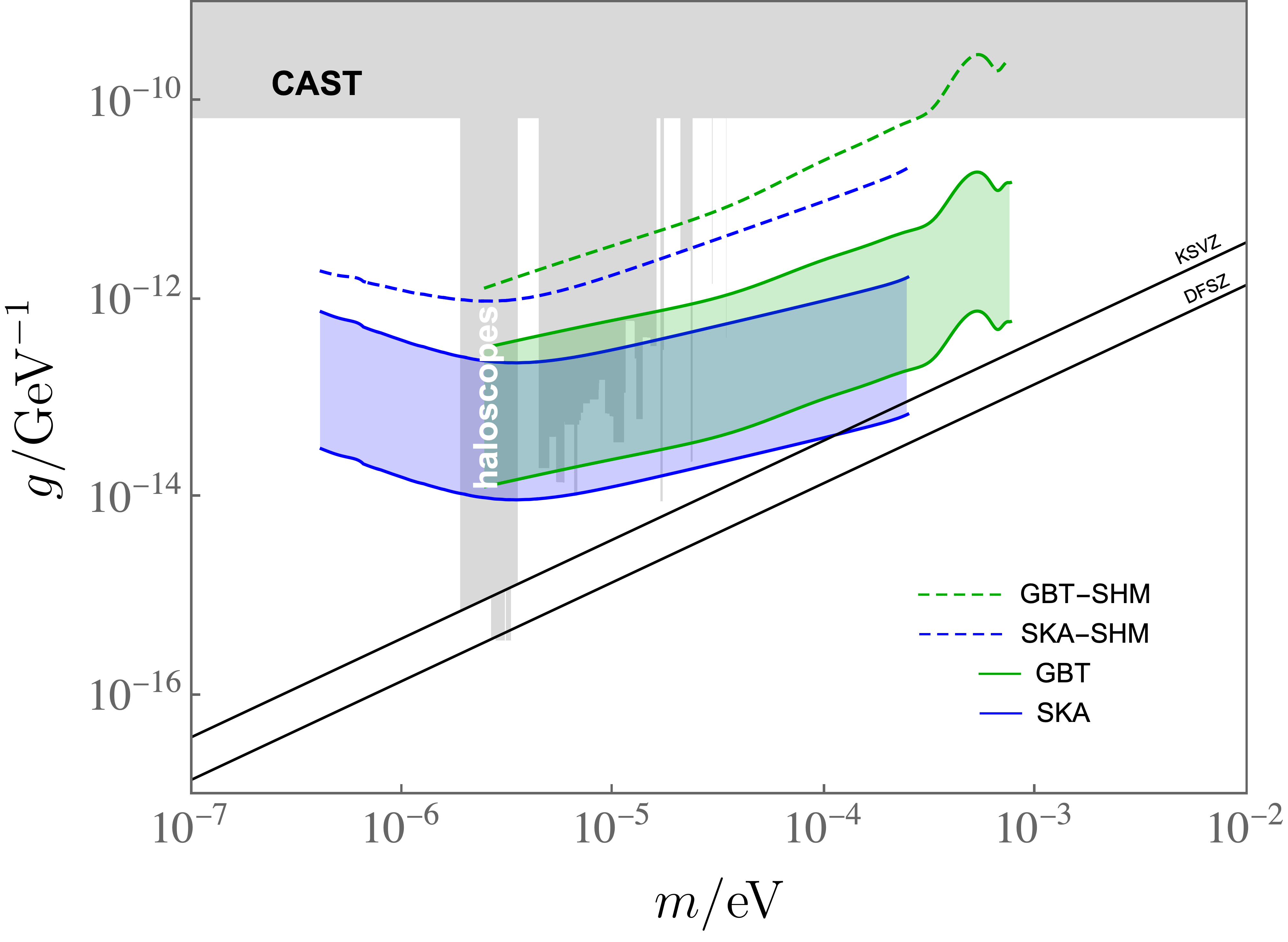}
\caption{Projected sensitivity for the low density stream scenario. The solid and dashed green (blue) lines are estimates for the GBT (SKA) in the case of low density streams and standard halo model, respectively. The widths of the sensitivity arises from the uncertainty in direction of the stream flow (see the text for more details). We have included current bounds from the CAST experiment \cite{Adair:2022rtw}, and resonant haloscopes as ADMX \cite{ADMX:2021nhd}, HAYSTAC \cite{HAYSTAC:2020kwv}, CAPP \cite{Lee:2022mnc}, and others \cite{CAST:2020rlf,Alesini:2022lnp,Quiskamp:2022pks}}
\label{fig:senslowstreams}
\end{figure}

So far our sensitivity for this scenario seems to have a higher potential because of the high energy density of the hairs combined with the small frequency width of the signal. However, it is dramatically reduced because of the lack of knowledge of the hairs location. Our approach is to fix the outgoing beam in a particular direction; in this work we choose the zenith because the beam can travel along the whole hair length and extract the maximum possible echo signal. We define $r$ as the frequency of appearance of hairs in the zenith direction. This is given in Tab. \ref{tab:daily_rate} for some sample values. This frequency depends on the hair density and also the volume of the sky under consideration. For instance, as the echo photons from hairs located at distances farther than $h_\text{min}$ (see Eq. \eqref{hmin1}) can not be collected, we are limited to consider hairs located in a spherical shell with inner and outer radii given by $R_E$ and $R_E+h_\text{min}$, respectively. Once the rate $r$ is given, the required frequency band of the beam is determined in order to cover a factor of two in axion mass range assuming a time interval $t_T=1\,\text{yr}$. In terms of $r$, the bandwidth of the outgoing photon beam must be

\beq
\Delta\nu={m\over4\pi r\,t_T}.
\eeq
On the other hand, the duration of the beam must be $\toff=1/r$.

Putting all together, the signal-to-noise ratio for this scenario becomes
\beq
s/n={1\over8}\sqrt{\pi\over2}{g^2\rho\,t_T r\over m\,T_n}{\cal T}_cP_0\sqrt{t_m\over B} \enspace, \label{sn3}
\eeq
where $t_m$ and $B$ are given by Eqs. \eqref{tm2} and \eqref{B1}, respectively.

\begin{figure}[t]
\includegraphics[width=1\linewidth]{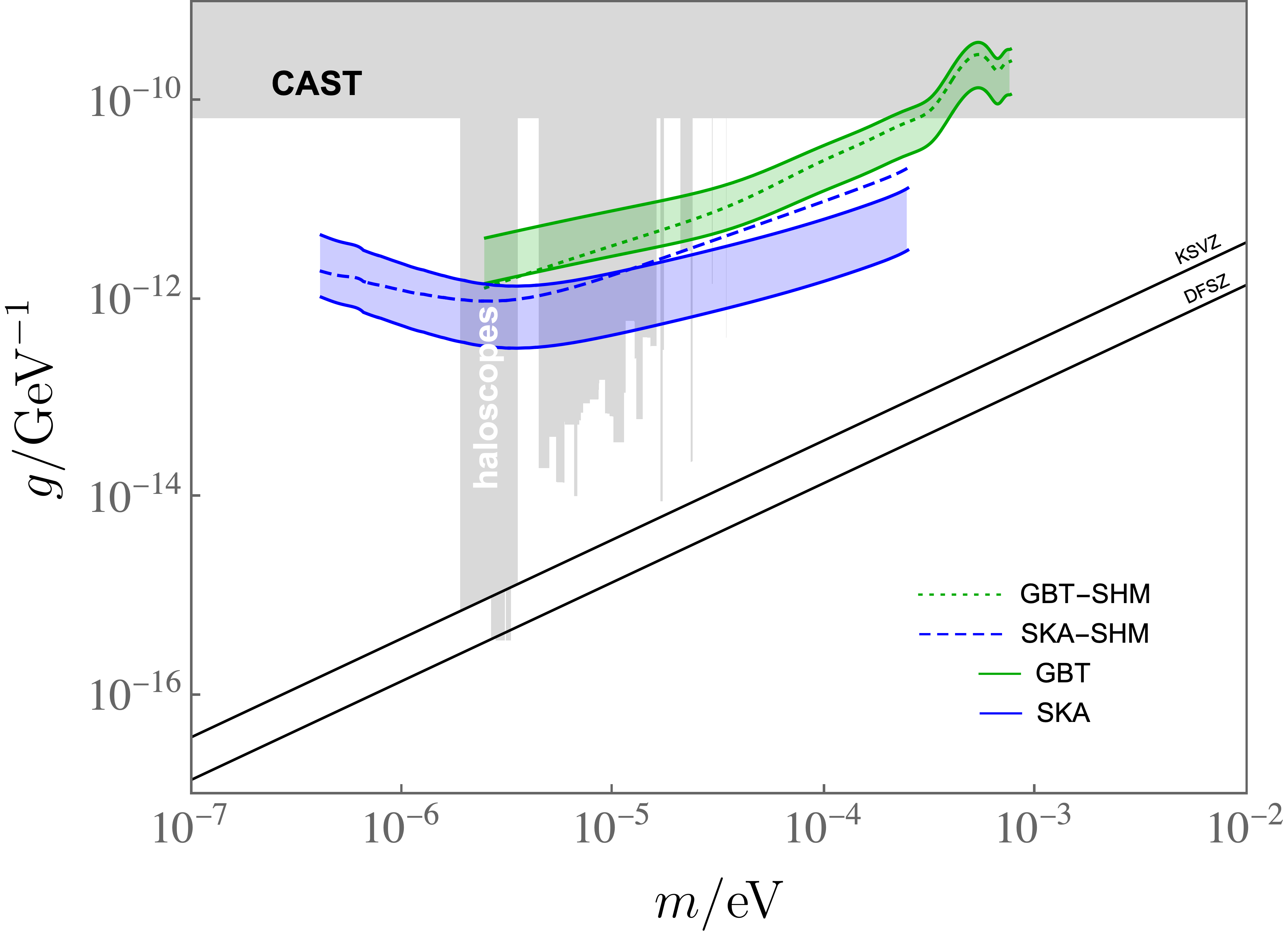}
\caption{Projected sensitivity for the gravitational hairs scenario. The solid and dashed green (blue) lines are estimates for the GBT (SKA) in the case of low density streams and standard halo model, respectively. The widths of the sensitivity arises from the uncertainty in location of the hairs (see the text for more details). We have included the same currents bounds as in Fig. \ref{fig:senslowstreams}.}
\label{fig:senshairs}
\end{figure}

\section{Results} \label{results}

For both scenarios we assume several experimental parameter values commonly used. For the sensitivity estimations we consider a total power of $P_0=10\,\text{MW}$ for the outgoing photon beam. We are aiming for covering a factor of two in axion mass range per year. In this work we take, as examples, two known radio telescopes to compute the projected sensitivity. First the single dish Green Bank telescope (GBT) \cite{GBT}, and then the Square Kilometre Array (SKA) \cite{SKA}. The radius of the GBT dish is $R_c=50\,\text{m}$ and its operation frequency band is from $0.3$ to $92\,\text{GHz}$. For the total noise temperature as a function of frequency, we use data from a typical measurement given in Ref. \cite{GBT}. Regarding SKA, it is aimed to work in two configurations; SKA1-low and SKA1-mid. SKA1-low consists of an array of 131,000 log-periodic dual-polarised antenna elements. Most of them are located in a very compact core of 500 m radius. Other antennae will be located out of the compact core, but for simplicity we only consider the compact core as an effective detector dish with radius $R_c=500\,\text{m}$. The working frequency range of SKA1-low is from 50 to 350 MHz with a stable receiver noise of 40 K. For the other SKA configuration the frequency range is from 350 MHz to 30 GHz, equipped with 133 dishes with 15 meter diameter and 64 dishes with 13.5 meters diameter. The dishes are distributed to form a moderately compact circle of radius $R_c=500\,\text{m}$. In both configurations we have the same collecting radius, so without going into detection details, we can estimate the sensitivity assuming the same physical parameters for the whole frequency band between 50 MHz and 30 GHz. For typical noise temperature for SKA, we use the one given in Ref. \cite{SKA}

\subsection{Low density axion streams}

As discussed in Sec. \ref{streams}, there is a huge number of fine grained streams with different energy density passing through.  Unfortunately, the probability that high density streams pass through the solar system is small. As shown in Table \ref{tab:stream_table}, the highest density streams accessible have density $\rho=10^{-3}\rho_0$. In fact, there is a 100\% probability to have 10 of them in our vicinity at any time.

As the stream direction is unknown, the value of $v_\perp$ at Eq. \eqref{sn2} ranges from $v_r\cos(\phi)$ (the velocity of the Earth rotation) to the stream velocity $v_s$. The former corresponds to the case when the outgoing photon beam is aligned with the stream and the latter when it is perpendicular to the stream. The projected sensitivity for this case is shown in Fig. \ref{fig:senslowstreams}, using Eq. \eqref{sn2} for $s/n=5$ and $\eta=10^{-7}$. The vertical width of the sensitivity regions comes from the uncertainty in $v_\perp$ mentioned above, where the lower bound gives the optimal sensitivity, i.e., for a beam aligned with the axion stream flow, and the upper bound is the sensitivity for the photon beam propagating perpendicular to the axion stream. For the aligned case we have $v_\perp=434\,\text{m/s}\cos(\phi)$, where $\phi$ is the latitude of the experiment. For the GBT we have $v_\perp\simeq434\,\text{m/s}\cos(38^o)=342\,\text{m/s}$ while for SKA we have $v_\perp\simeq434\,\text{m/s}\cos(30^o)=376\,\text{m/s}$. If the outgoing beam is perpendicular to the axion stream, the value of $v_\perp$ is taken to be $220\,\text{km/s}$ for a conservative estimate.

\subsection{Axion hairs}

For gravitational hairs~\cite{Prezeau:2015, gfstream:2022} we first notice that the length of the hairs of about $0.2 R_E\simeq1.28\times10^3\,\text{km}$ is smaller than the parameter $R_c/v_\perp$ for both radio telescope examples used here. Indeed, using $v_\perp=v_r\cos(\phi)$ we have $R_c/v_\perp= 4.37\times10^4\,\text{km}$ for the GBT and $R_c/v_\perp=3.98\times10^5\,\text{km}$ for SKA. Then we will take $l\simeq 0.2R_E$ for the minimum value expressed in Eq. \eqref{Tc1}.

As discussed in Sec. \ref{echo signal}, we can receive axion echos from hairs located not farther than $h_\text{min}$ given by Eq. \eqref{hmin1}. For the GBT ($R_c=50\,\text{m}$ and $v_\perp=342\,\text{m/s}$) $h_\text{min}=3.43R_E$ while for SKA ($R_c=500\,\text{m}$ and $v_\perp=376\,\text{m/s}$) $h_\text{min}=31.2R_E$. From Tab. \ref{tab:daily_rate} we use values for $r$ and $\rho$ according to places within $3R_E$ spherical shell for the GBT, and $20R_E$ for the SKA. As we can see in Eq. \eqref{sn3}, the detection sensitivity is proportional to the product $\rho\cdot r$. Then the most promising target, among the ones listed in Tab. \ref{tab:daily_rate}, are hairs with density $\rho=10\,\rho_0$ and whose rate of appearance are $r=0.05/\text{day}$ for the GBT and $r=0.4/\text{day}$ for SKA. According to the discussion at the end of Sec. \ref{echo signal hairs}, for the GBT we assume 18 photon shots, each one lasting for $\toff= 20\,\text{days}$ and with a bandwidth of $\Delta\omega=416\,\text{MHz}\left(m\over10^{-5}\text{eV}\right)$, during one year, to cover a factor of two in axion mass range. For SKA we need 146 shots, each one with duration $\toff=2.5\,\text{days}$ and bandwidth $\Delta\omega=52\,\text{MHz}\left(m\over10^{-5}\text{eV}\right)$.

The detection sensitivity depends strongly on the observation time given by Eq. \eqref{tm2}. This becomes optimal if the hair is at the Earth surface, which occurs for stream velocities around the critical value $v_s=11.2\,\text{km/s}$. The worst case corresponds to a hair located just at our observational limit, defined by Eq. \eqref{hmin1}. The worst case for the GBT ($h_\text{min}=3R_E$) corresponds to an axion stream velocity of $v_s=19.4\,\text{km/s}$, while for SKA ($h_\text{min}=20R_E$) the corresponding stream velocity is $v_s=50.1\,\text{km/s}$. With respect to the diameter $D$ of the hairs, it is shown to maintain a uniform value of $\sim 5\,\text{km}$, regardless of the hair location~\cite{gfstream:2022}.

The projected sensitivity for axion gravitational hairs is shown in Fig. \ref{fig:senshairs} for the GBT and SKA. The upper and lower sensitivity bounds are due to the unpredictable location of the hairs, as discussed above.

\section{Conclusion} \label{conclusion}

In this work we use the echo method \cite{Arza:2019nta} to explore the possibility to search for DM axions in the form of cosmologically motivated fine grained streams. Of note, the fine grained streams are remnant flows from the early stages of the galactic structure formation~\cite{Stucker:2020,Vogelsberger:2020}, with very small velocity dispersion. Two scenarios are considered. First, low density streams, likely to be present in the solar system, and second, high density hairs formed by gravitational self focusing effects by the inner Earth. It is shown that for both cases the performance of the echo method is notably improved with respect to the virialized SHM.  The case of the low density streams is the more favorable despite the low dark matter axion energy density. The main reason for this is the small probability to encounter a hair by the photon beam, leading to an unavoidable loss in sensitivity. 

The echo method is a new concept for axion dark matter search based on detecting electromagnetic echoes produced from stimulated axion decay. Its sensitivity depends substantially on the local dark matter phase space distribution. Even though, it is shown to be a promising technique when considering the SHM~\cite{Arza:2019nta}, its efficiency can be largely improved for DM axion scenarios with very cold flows. Indeed, it was already demonstrated for the caustic ring model in Refs. \cite{Arza:2019nta,Arza:2021nec}, and here it is also shown for the case of fine grained axion streams in the solar system, where the improvement in sensitivity can be as large as two orders of magnitude. 

\acknowledgments

K.Z. is grateful to Pierre Sikivie for triggering his interest in the echo idea during a seminar he gave at CERN. A.A. thanks the Institute for Theoretical and Mathematical Physics (ITMP) of Lomonosov Moscow State University for their support during difficult times. AK acknowledges Messiah University Scholarship Programs for support.

\bibliography{biblio.bib}

\end{document}